\documentclass[10pt,a4paper]{article}
\usepackage{jheppub_kim}
\usepackage{pdflscape}
\usepackage{amsmath}
\usepackage{amssymb}
\usepackage{dcolumn}
\usepackage{bm}
\usepackage{color}
\usepackage{epsfig}
\usepackage{amsfonts}
\usepackage{graphicx}
\usepackage{subfigure}
\UseRawInputEncoding
\begin{document}
\title{Unveiling the Fifth State of Matter: Insights into Ultra-Hot Plasma and its Applications}
\author[a]{Mohammad Mehdi Bagheri-Mohagheghi,}
\author[a,b,c,d]{Behnam Pourhassan,}
\author[e,f,g]{Emmanuel Saridakis,}
\author[h,i]{Salvatore Capozziello,}
\author[j]{Prabir Rudra}

\affiliation[a] {School of Physics, Damghan University, Damghan, 3671641167, Iran.}
\affiliation[b] {Physics Department, Istanbul Technical University, Istanbul 34469, Turkey.}
\affiliation[c] {Canadian Quantum Research Center 204-3002 32 Ave Vernon, BC V1T 2L7 Canada.}
\affiliation[d] {University of British Columbia Okanagan, Kelowna, Canada.}
\affiliation[e] {Institute for Astronomy, Astrophysics, Space Applications and Remote Sensing (IAASARS), National Observatory of Athens,
Athens, Greece.}
\affiliation[f]{Deep Space Exploration Laboratory/School of Physical Sciences,
University of Science and
Technology of China, Hefei, Anhui 230026, China.}
\affiliation[g] {Departamento de Matem\'{a}ticas, Universidad Cat\'{o}lica del
Norte, Avda. Angamos 0610, Casilla 1280 Antofagasta, Chile.}

\affiliation[h] {Dipartimento di Fisica, "E. Pancini", Universita' "Federico

II" di Napoli, Compl. Univ. Monte S. Angelo Ed. G, Via Cinthia, I-80126 Napoli,
Italy.}
\affiliation[i] {Scuola Superiore Meridionale, Largo S. Marcellino 10, I-80138
Napoli, Italy.}
\affiliation[j] {Department of Mathematics, Asutosh College, Kolkata-700 026,
India.}

\emailAdd{bmohagheghi@du.ac.ir}
\emailAdd{b.pourhassan@du.ac.ir}
\emailAdd{msaridak@noa.gr}
\emailAdd{capozziello@unina.it}
\emailAdd{prudra.math@gmail.com}

\abstract{In this work we investigate the dissociation energy of the
North (N) and South (S) poles of a quantum magnetic particle, incorporated
within both classical and quantum mechanical perspectives. A simple model of a
harmonic oscillator is employed to estimate the dissociation energy of the N-S
poles, as well as the corresponding breakdown temperature and internal pressure.
The results indicate that the separation of magnetic poles occurs in two states:
(a) in an ultra-hot plasma medium with extremely high temperatures, such as in
the core of a hot star, and (b) at extremely high pressures, such as between
internal plates in complex superlattices of layered solids. The breakdown
temperature is found to be of the order of $10^7$ to $10^8$ Kelvin, which is
only achievable in an ultra-hot plasma environment, known as the fifth phase of
matter. Based on this model, the possibility of dissociation of bonds between N
and S magnetic poles for solid superlattices under very high pressures between
crystal plates is also calculated. The results suggest that the presence of
isolated magnetic monopoles in superlattices of solids under
ultra-high-pressure conditions is possible. Consequently, the model proposes
that the conductivity of magnetic monopole carriers can be applied in the
manipulation of nanomaterials for the production of advanced devices, such as
new generations of superconductors, new spin devices and magnetic-electronics,
advanced materials with magnetic monopoles, as well as super-dielectrics.}

\maketitle

\section{Introduction}\label{I}

\begin{figure}[h!]
\begin{center}$
\begin{array}{cccc}
\includegraphics[width=60 mm]{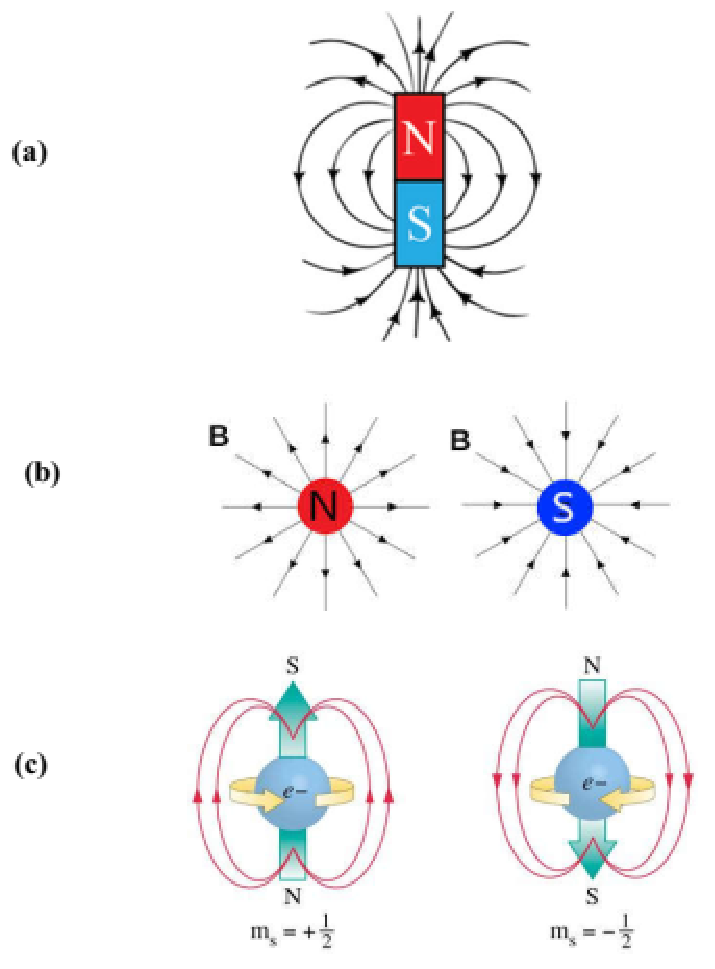}
\end{array}$
\end{center}
\caption{{\it{(a) Magnetic dipoles and magnetic field of (B) around it
and (b) Magnetic monopoles with isolated separated monopoles of S
and N and magnetic field of (B) around it and (c) magnetic moments
of an electron in the opposite spin directions \cite{1}.}}}
\label{fig1}
\end{figure}

New physics could arise from interactions at particle colliders, primarily
depending on the new symmetries   at high energies or in extra
dimensions.  This could bring about  fundamental changes in our understanding
of the properties during the early universe, due to the presence of new
particles in the primordial thermal bath or through phase transitions.
Magnetic matter  is characterized by two magnetic poles, namely N and S, even on
a very small atomic scale. The existence of magnetic dipoles in normal
conditions is attributed to a fundamental property of magnetic materials that
arises from electron spin. In the standard mode of particle physics  it is not
possible to separate the two magnetic poles. However, in more complicated
theories one may have the appearance of magnetic monopoles
\cite{tHooft:1974kcl,Montonen:1977sn,Preskill:1979zi,Gross:1983hb,
Mavromatos:2016mnj,Mavromatos:2020gwk,Addazi:2021xuf,Luciano:2023bai}, while
recent advanced  experiments conducted at very high pressures have started
lighting the issue  \cite{1, MoEDAL:2016lxh,MoEDAL:2017vhz}. In Fig. \ref{fig1}
we demonstrate that magnetic monopoles behaving like single electric charges,
with positive and negative polarity, creating an inward or outward magnetic
field around themselves \cite{1, MoEDAL:2016lxh,MoEDAL:2017vhz,2,3}.

There has been  a growing interest in the study of magnetic monopoles, with
international experiments being conducted to detect them in artificial spin ice
systems \cite{3,4,5}. Artificial spin ice materials have attracted theoretical
and experimental attention in recent years \cite{5,6,7,8,9}, as they are related
to magnetic charge defects possessing similar properties to those of magnetic
monopoles postulated by Dirac. In particular, the ``Duality in nature
symmetry'' provides both electric and magnetic field sources, including magnetic
monopoles and magnetic currents. Although magnetic monopoles are elusive as
elementary particles, they exist in many materials in the form of quasiparticle
elementary excitations. On the other hand, experiments have directly measured
magnetic monopoles and associated currents, thereby confirming the predicted
symmetry between electricity and magnetism. This field encompasses the
conditions for the formation of magnetic monopoles, electrical and thermal
conductivity, and their wave and energy equations \cite{3,9}. Finally, symmetry
in nature is an important subject in this field, predicting the existence of
itinerant magnetic monopoles as single free charges at low temperatures in
Maxwell's equations.

Dirac's theory  suggests that the existence of magnetic monopoles with an
electric unit charge of $Q= (137) e$ is possible, as per the $eg = n\hbar
c/2$ Dirac quantization condition (DQC), due to the convergence of nature  and
the principles of physical phenomenology \cite{11,12,13}. However, it is not
possible to create a magnetic monopole under normal conditions. Only with very
high energy, mechanical forces, or ultra-high pressures can magnetic dipoles be
broken down into two free magnetic monopoles. Furthermore, applying external
pressures in the Giga-Pascal range can induce thermodynamic changes and create
new phenomena in complex solid-state structures under critical conditions. For
instance, a pressure of a few Giga-pascals can lead to very good conductivity in
crystalline helium, as well as in diamonds under high pressures. Induced
pressure can alter the initial nature and physical properties of matter
\cite{14,15,16,17}.

In recent years,  significant research has been conducted on spin-ice, revealing
the existence of magnetic monopoles in some superlattice structures of solids
such as pyrochlores structures and superconductor materials. This results in a
bound pair of north and south poles that can be fractionalized into two free
magnetic monopoles \cite{17,18,19,20,21}. Presently, the spin-ice properties of
pyrochlore magnetic materials, including ${Yb}{2}{Ti}{2}O_7$,
${Er}{2}{Ti}{2}O_7$, and ${Dy}{2}{Sn}{2}O_7$, have been investigated. These
materials host strong quantum fluctuations of magnetic dipoles due to pseudospin
(1/2) in magnetic rare-earth elements \cite{22,23,24}.

Thouless et al.  (winners of the Physics Nobel Prize in 2016) recently developed
a topological model for studying critical phenomena and phase transitions.
According to this theory, phase transitions of matter are associated with
topological defects, such as vortices or structural holes. The model
demonstrates topological phase changes in critical phenomena, such as
superconductivity, magnetic ultra-thin films, and superfluids, caused by
topological defects \cite{25,26}. Magnetic monopoles can be considered
topological holes based on this theory. Research on magnetic monopoles has
expanded to various areas of theoretical and experimental physics, including
elementary particle physics, solid-state physics, astrophysics, gravity, and
cosmology \cite{27,28,29,30}.

In the present work we  present a simple mechanical and quantum
model, based on the energy of a simple harmonic oscillator, to investigate the
breaking conditions of magnetic dipole junctions and to obtain their
corresponding temperature and internal pressure in solids. The temperature
required for this phenomenon is very high, and it can only occur in an ultra-hot
plasma environment, which constitutes the fifth phase of matter. Additionally,
we will explore whether these conditions can also be achieved under high
mechanical pressures in quantum structures and superlattices of layered solids,
potentially offering numerous applications in the future.

\section{Calculation of dissociation energy of N-S poles}\label{II}

In this section we present the calculation of dissociation energy of N-S poles,
following both  a classical as well as a quantum approach.

\subsection{Theoretical model with classical approach}

\begin{figure}[h!]
\begin{center}$
\begin{array}{cccc}
\includegraphics[width=60 mm]{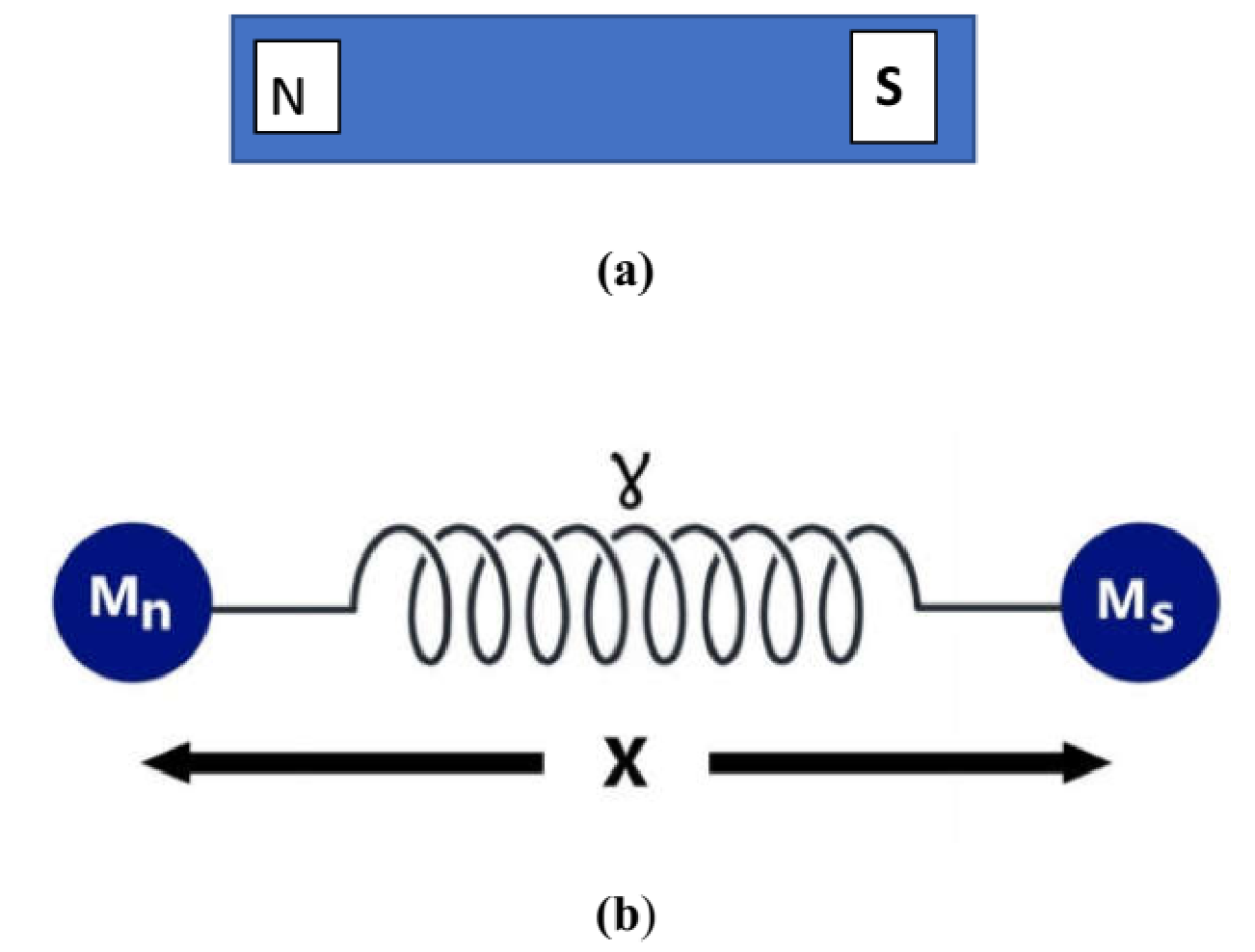}
\end{array}$
\end{center}
\caption{{\it{(a) A single magnetic  particle with N-S poles. (b) A two-mass
oscillator with $M=M_N=M_S = 137 m_e$ with relative distance of $x$.}}}
\label{fig2}
\end{figure}

Let us consider a magnetic  particle with N-S magnetic poles. These poles can be
modeled as a simple classic harmonic system, comprising two bodies of mass $M$
connected by a spring of elasticity $\gamma$ and a relative distance of $x$, as
illustrated in Fig. \ref{fig2}. By adopting this approach, it is possible to
compute the dissociation (binding) energy of N-S poles, which is analogous to a
classical physics model. Figs. \ref{fig2} (a) and (b) present a schematic view
of N-S magnetic poles of a magnetic particle that transform into two isolated
bodies, namely N-pole and S-pole.

Let assume that we have a two-body oscillator consisting of $M_N$ and
$M_S$ that can be represented as a single-body oscillator with reduced mass
($1/\mu = 1/M_N+1/M_S$). If $x$ denotes the relative change in the spring length
due to the force $F$, and we consider Newton's second law, we can obtain the
following differential equation for the distance ($x$) of the harmonic system:
\begin{equation}\label{1}
\mu\frac{d^{2}x}{dt^2}+\gamma x=0,
\end{equation}
where $\gamma$ is the elasticity of the spring. The solution to equation (\ref{1}) is given by
\begin{equation}\label{2}
x(t)=A\sin{\omega t},
\end{equation}
where $\omega$ is defined as
\begin{equation}\label{3}
\omega=\sqrt{\frac{\gamma}{\mu}}.
\end{equation}

\begin{figure}[h!]
\begin{center}$
\begin{array}{cccc}
\includegraphics[width=100 mm]{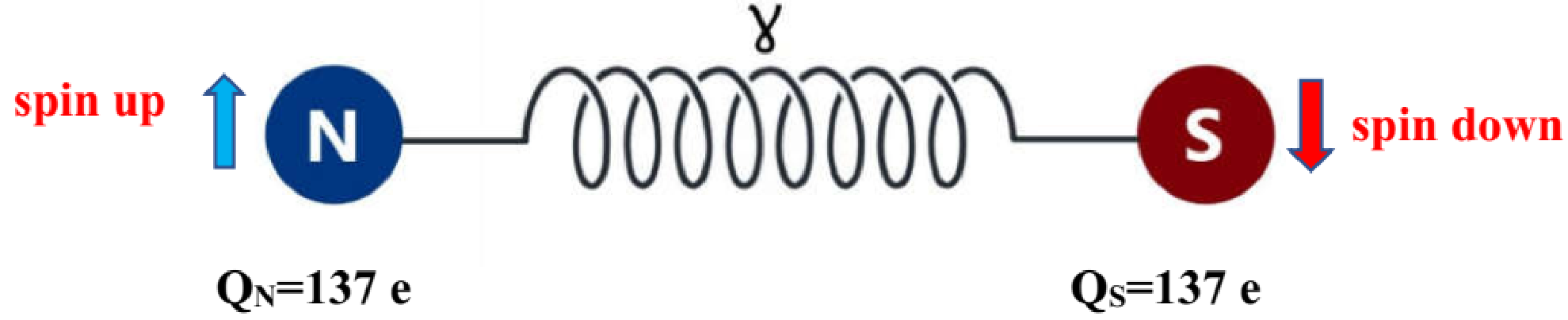}
\end{array}$
\end{center}
\caption{{\it{Segregated magnetic poles containing: pole N with 137
electrons of (mean) spin up $\uparrow$ and pole S with 137
electrons of (mean) spin down $\downarrow$.}}} \label{fig3}
\end{figure}

In order to calculate the oscillation frequency $\omega$ or the bonding energy
of dipoles, we require the spring elasticity constant ($\gamma$). The bonding
energy of magnetic poles is believed to have a relationship similar to the
Coulomb interaction force ($F_{N-S}$) between two electric charges with a
magnitude of $Q_N=Q_S (=137e)$ that have up and down spins. As predicted by
Dirac \cite{12,13} for a magnetic monopole based on quantum calculations, and
it is illustrated in Figure \ref{fig3}, it can be expressed as
\begin{equation}\label{4}
F_{N-S}=K\frac{Q_{N}Q_{S}}{x^{2}},
\end{equation}
where $K$ is the Coulomb  constant of the environment  which is equal to
$1/4\pi \varepsilon$. Additionally, $x$ is the relative distance between
magnetic monopoles, which  is in the sub-atomic range, specifically of the
order of the electron's effective radius, and here we consider it to be
$10^{-12} m$.

In Fig.  \ref{fig3} we illustrate a magnetic quantum particle with a single
magnetic charge of $Q_N$ and $Q_S$ equal to $137e$, where $\uparrow$ denotes the
spin-up state for the N pole and $\downarrow$ denotes the spin-down state for
the S pole. According to the classical mechanics equation of the simple harmonic
oscillator, the Coulomb force is equivalent to the elastic force of the
oscillator with an elastic constant $\gamma$, as expressed in the following
equation:
\begin{equation}\label{5}
F=K\frac{Q_{N}Q_{S}}{x^{2}}=-\gamma x.
\end{equation}
The elastic constant of the spring can be computed using equation
(\ref{5}), and the total energy of the system can be determined by considering
an equal contribution of potential and kinetic energy, namely
\begin{equation}\label{6}
\gamma=K\frac{Q_{N}Q_{S}}{x^{3}},
\end{equation}
and
\begin{equation}\label{7}
E=E_{k}+U=2U=2(\frac{1}{2}\gamma x^{2})=\gamma x^{2}.
\end{equation}
Substituting $\gamma$ from equation (\ref{6}) into the energy equation
(\ref{7}), we obtain the total energy of the magnetic bipolar as:
\begin{equation}\label{8}
E=K\frac{Q_{N}Q_{S}}{x^{3}}x^{2}=K\frac{Q_{N}Q_{S}}{x}.
\end{equation}

As an example, for this approximation, the dissociation energy of
the magnetic dipole in quantum scale is calculated by replacing the
approximate value of $x$ in equation (\ref{8}). Assuming  that a dielectric
constant $k = 10^3$ is given for a typical dielectric material, such as lead
zirconate titanate (refer to Table \ref{tab1}), we can calculate $K$ as
$K=1/(4\pi k \varepsilon_0 )= 10^{-3}\times(9\times10^9)$, where $x$ is the
relative change in spring length. By substituting $Q_N$ and $Q_S$ into equation
(\ref{8}), we can determine the total energy in two states:

(a) If $x$ is considered as $10^{-12} m$ (about ten times the effective electron
radius), the energy ($E$) can be estimated to be $2.7\times10^4$ electron volts
(eV), which is equivalent to the thermal energy at a temperature of
$2\times10^8$ Kelvin, such as that found in the core of hot stars. This energy
value can be expressed as:
\begin{equation}\label{10}
E=\frac{3}{2}k_{B}\theta_{B},
\end{equation}
where $k_{B}$ is the Boltzmann constant. By substituting the values of $E$ and
$k_{B}$, we can obtain $\theta_{B}=2\times10^8$ Kelvin.

\begin{table}[tbp]
\centering
\begin{tabular}{|c|c|c|c|c|}
\hline
      &Material          & Dielectric constant ($\varepsilon_r$)    \\
\hline
\hline
1     &Pyrex (Glass)                & 4.7 (3.7-10)                             \\
\hline
2     &Diamond                      &5.5-10                                    \\
\hline
3     &Graphite                     &10-15                                     \\
\hline
4     &Silicon                      &11.68                                     \\
\hline
5     &Glycerol                     &42.5(at 25 °C)                            \\
\hline
6     &Titanium dioxide             &86-173                                    \\
\hline
7     &Strontium titanate           &310                                       \\
\hline
8     &Barium strontium titanate    &500                                       \\
\hline
9     &Barium titanate              &1200-10,000(20-120 °C)                    \\
\hline
10    &Lead zirconate titanate      &500-6000                                  \\
\hline
11    &Conjugated polymers          &1.8-6 up to 100,000                       \\
\hline
12    &Calcium copper titanate      &$>250$                                    \\
\hline
\end{tabular}
\caption{\label{tab1} Relative permittivity of various materials at room
temperature \cite{31}.}
\end{table}

(b) If $x$ is considered  as $10^{-9} m$ (1nm), such as in the case of a
magnetic quantum dot, the corresponding temperature for equivalent thermal
energy will be on the order of 105 Kelvin. However, this result is in conflict
with the actual corresponding temperature, which is on the order of 108 Kelvin.
Hence, this is an indication  that the equivalent energy is
highly dependent on the mean distance between magnetic dipoles.

The mean distance between magnetic dipoles  in a given material can be
determined through various experimental techniques, depending on the specific
properties of the material and the type of magnetic interaction involved. One
common technique is to measure the magnetic susceptibility of the material,
which is the extent to which the material becomes magnetized in response to an
applied magnetic field. This can provide information about the magnetic
properties of the material and the degree of magnetic ordering or clustering of
the dipoles.

Another technique is to use  microscopy methods such as scanning electron
microscopy (SEM) or transmission electron microscopy (TEM) to directly image the
magnetic structure of the material at the nanoscale level. This can give insight
into the spatial arrangement and distribution of the magnetic dipoles.
Furthermore, other methods may involve measuring the magnetic resonance
properties of the material, such as the nuclear magnetic resonance (NMR) or the
electron paramagnetic resonance (EPR), which can offer information about the
dynamics and interactions of the magnetic dipoles.

We should mention here that theoretical modeling  and experimental measurements
collaborate in order to determine the mean distance between magnetic dipoles in
a material, by providing complementary information about the physical
properties of the system. Theoretical models can provide a framework for
understanding the underlying physical mechanisms that govern the magnetic
interactions between dipoles, and can predict the expected behavior of the
system under different conditions. These models often involve simplifying
assumptions and approximations, and may require input parameters based on
experimental measurements, such as magnetic susceptibility or magnetic resonance
properties. On the other hand, experimental measurements provide direct
observation of the physical properties of the material, such as the magnetic
susceptibility or the spatial arrangement of the dipoles. These measurements can
validate or refine the assumptions and parameters used in   theoretical
models, and can offer insights into the limitations of the model. By combining
the information from theoretical models and experimental measurements,
researchers can refine their understanding of the magnetic properties of the
material and improve their predictions of the mean distance between magnetic
dipoles. This iterative process of modeling and experimentation is a key aspect
of scientific research, and can lead to deeper insights into the fundamental
physical processes that govern the behavior of magnetic systems.

\subsection{Theoretical model with quantum approach}

A quantum approach is also possible by considering the system as a quantum
harmonic oscillator. For a quantum oscillator with a reduced mass of $\mu$  we
have
\begin{equation}\label{11}
E=(n+\frac{1}{2})\hbar\omega,
\end{equation}
where $\omega=\sqrt{\frac{\gamma}{\mu}}$, while the minimum energy (base
state) of the oscillator, corresponding to $n = 0$, can be expressed as:
\begin{equation}\label{12}
E_{min}=\frac{1}{2}\hbar\omega.
\end{equation}
By substituting $\gamma$ from equation (\ref{6}) into equation (\ref{11}) for
$\omega$, and considering the reduced mass $\mu = M/2$ with $M = 137m_e$, we
obtain:
\begin{equation}\label{14}
E=\frac{\hbar}{2}\sqrt{\frac{2KQ_{N}Q_{S}}{Mx^{3}}}.
\end{equation}

Equation (\ref{14}) expresses the total energy of the magnetic bipolar system
in terms of the quantum harmonic oscillator model, using the reduced mass $\mu$
and the angular frequency $\omega$ of the oscillator. Additionally, it shows
that the energy of the system is proportional to the square root of the product
of the spring constant $K$, the magnetic charges $Q_N$ and $Q_S$, and the
inverse cube of the relative change in spring length $x$.

In the context of the quantum harmonic oscillator, equation (\ref{14}) provides
a way to calculate the energy levels of the system based on the quantum number
$n$, which determines the number of energy quanta or ``oscillations'' in the
system. The equation shows that the minimum energy of the system is proportional
to the angular frequency $\omega$, which depends on the spring constant and the
reduced mass.

The significance of equation (\ref{14}) in the context of the quantum harmonic
oscillator is that it provides a theoretical framework for understanding the
behavior of the magnetic bipolar system in terms of quantum mechanical
principles. By using this equation, one can make predictions about the
energy levels and properties of the system, and compare these predictions with
experimental measurements to test the validity of the model.

Suppose we have a magnetic bipolar system with magnetic charges $Q_N = Q_S =
5\times 10^{-18} \text{C}$, a spring constant $K = 2\times10^{-6} \text{N/m}$,
and a relative change in spring length $x = 10^{-9} \text{m}$. We can calculate
the reduced mass of the system using $\mu = M/2$, with $M = 137m_e$, where $m_e$
is the mass of the electron, acquiring
\begin{equation}
\mu = \frac{M}{2} = \frac{137m_e}{2} \approx 1.23\times 10^{-26} \text{kg}.
\end{equation}
We can then calculate the  angular frequency of the quantum harmonic oscillator
using the expression $\omega = \sqrt{\frac{\gamma}{\mu}}$, where $\gamma =
KQ_NQ_S/x^3$, which  yields
\begin{equation}
\omega = \sqrt{\frac{\gamma}{\mu}} =  \sqrt{\frac{2KQ_NQ_S}{Mx^3}} \approx
1.05\times 10^{16} \text{rad/s}.
\end{equation}
 Using (\ref{11}) we can calculate the energy levels of the system for
different values of the quantum number $n$. For example, when $n=0$ we can use
  (\ref{12}) to calculate the minimum energy of the system as
\begin{equation}
E_{min} = \frac{1}{2}\hbar\omega \approx 3.29\times 10^{-20} \text{J}.
\end{equation}
Similarly, when $n=1$ the energy of the system can be calculated using equation
(\ref{11}) as
\begin{equation}
E = \left(1 + \frac{1}{2}\right)\hbar\omega \approx 9.88\times 10^{-20}
\text{J},
\end{equation}
while for $n=2$ the energy is
\begin{equation}
E = \left(2 + \frac{1}{2}\right)\hbar\omega \approx 1.65\times 10^{-19} \text{J}.
\end{equation}
By calculating the energy levels of the system for different values of $n$, we
can gain insight into the behavior and properties of the magnetic bipolar
system, and compare these predictions with experimental measurements to test the
validity of the model.

In addition to  equation (\ref{11}) we have the
Schr$\ddot{\text{o}}$dinger equation, a fundamental equation of
quantum mechanics that describes the behavior of a system in terms of its wave
function. The time-independent Schr$\ddot{\text{o}}$dinger  equation for a
one-dimensional harmonic oscillator is given by:
\begin{equation}
\frac{d^2\psi(x)}{dx^2} + \frac{2m}{\hbar^2} \left[E -  \frac{1}{2}m\omega^2
x^2\right]\psi(x) = 0,
\end{equation}
where $\psi(x)$ is the system wave function, $m$ is the mass of the
particle, $\omega$ is the angular frequency of the oscillator, and $E$ is the
total energy of the system.
The solutions to this equation are the  energy eigenstates of the system, which
are characterized by discrete energy levels that are quantized in units of
$\hbar\omega$. The energy levels are
\begin{equation}
E_n = \left(n+\frac{1}{2}\right)\hbar\omega,
\end{equation}
where $n$ is the quantum number that determines the number of energy  quanta or
``oscillations'' in the system. The Schr$\ddot{\text{o}}$dinger equation
provides a more rigorous and general approach to calculating the energy levels,
and can be used to study the behavior of the magnetic bipolar system in more
detail. However, it can be more complex and computationally demanding than the
simpler harmonic oscillator model expressed by equation (\ref{11}).

Based on the above  discussion, we can estimate the energy of a quantum
oscillator using $Q_N = Q_S = 137 e$ and a mean distance of $x = 10^{-12} m$. In
this case, we can use   (\ref{14}) to obtain the energy of the system,
which is approximately
\begin{equation}\label{15}
E=4.38\times 10^{-16}J=2.74\times 10^{3} eV.
\end{equation}
Furthermore, we can calculate the equivalent temperature of the system
using the Boltzmann constant $k_B$ and the energy $E$, obtaining
\begin{equation}\label{16}
\theta=\frac{E}{k_B}=2.1\times 10^{7} K.
\end{equation}
These expressions  provide a way to estimate the energy and temperature of a
quantum oscillator based on the magnetic charges and mean distance between the
magnetic dipoles. These estimates can be used to gain insight into the behavior
and properties of the system, and can be compared with experimental measurements
to test the validity of the model. Note that it is possible to calculate the
energy and temperature of a system with different magnetic charges using
(\ref{14}) and the Boltzmann constant, as long as the inter-dipole separation
$x$ and the spring constant $K$ are known. To calculate the energy of the
system, we can simply substitute the new values of the magnetic charges $Q_N$
and $Q_S$ into   (\ref{14}), along with the known values of $x$ and $K$,
and solve for $E$. Similarly, we can calculate the equivalent temperature of the
system using the Boltzmann constant $k_B$ and the energy $E$, as given by
  (\ref{16}). It is important to mention that  relation (\ref{14}) assumes
that the system is a quantum harmonic oscillator, and that the magnetic dipoles
are in thermal equilibrium with their surroundings. In reality, there may be
other factors that affect the energy and temperature of the system, such as
external magnetic fields, thermal gradients, or non-equilibrium initial
conditions. Therefore, the calculated values of energy and temperature should be
interpreted as estimations, and be compared with experimental measurements to
test the validity of the assumptions and model.

\begin{figure}[h!]
\begin{center}$
\begin{array}{cccc}
\includegraphics[width=100 mm]{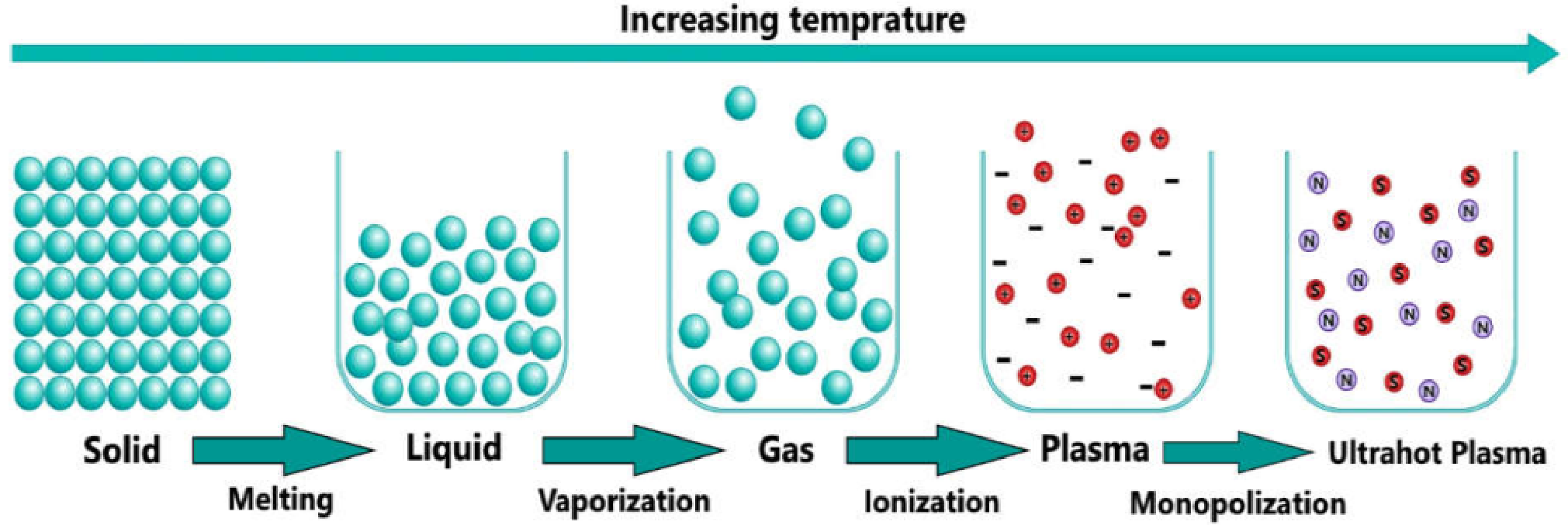}
\end{array}$
\end{center}
\caption{{\it{The phase change of matter from solid to ultra-hot plasma
with increasing temperature.} }}\label{fig4}
\end{figure}

According to the quantum  harmonic oscillator model, the breakdown temperature
$\theta$ is of the order of $10^7$, which agrees reasonably well with the
results obtained from the mechanical approach. However, it is important to note
that under normal conditions, it is not possible for the magnetic poles to
separate, and this only occurs in extremely hot environments such as hot stars
where matter exists in the ultra-hot plasma phase. In this phase, matter becomes
suitably hot that it undergoes ionization and the magnetic poles separate,
resulting to the existence of magnetic monopoles. This state of matter is
referred to as the fifth state of matter. Fig. \ref{fig4} illustrates the
phase change from the solid state to the ultra-hot plasma phase,  as
temperature increases.

The investigation of the fifth state of matter can be performed in the
following fields:

\textbf{Nuclear fusion:} The ultra-hot plasma phase is relevant to the study of
nuclear fusion, which involves the combination of atomic nuclei to form heavier
elements with the release of energy. In order to achieve nuclear fusion, a
plasma of hydrogen isotopes must be heated to ultra-high temperatures and
densities, which can be achieved by using magnetic confinement techniques.
Understanding the behavior of matter in the ultra-hot plasma phase is essential
for the development of practical fusion reactors.

\textbf{Astrophysics:} The fifth state of  matter is   relevant to the study
of astrophysics, as it is present in high-energy environments such as stars,
supernovae, and active galactic nuclei. Understanding the behavior of ultra-hot
plasmas in these environments is essential for understanding the dynamics and
evolution of the universe.

\textbf{Materials science:} The fifth state of matter is relevant to the study
of materials science, as it can be used to produce materials with unique
properties and applications. For example, ultra-hot plasma can be used to create
thin films, nanoparticles, and other materials with precise control   their
properties, such as composition, size, and structure.

\textbf{Plasma processing:}  The fifth state of matter is   used in plasma
processing, which involves using plasma to modify the surface properties of
materials. For instance, the plasma can be used to clean,  , or deposit
materials with high precision, which is useful in the fabrication of
microelectronics, solar cells, and other devices.

In summary, understanding the fifth state of matter is important for advancing
our knowledge and capabilities in a wide range of fields, from energy production
and astrophysics to materials science and microelectronics.

\section{Calculation of dissociation pressure of N-S poles in solids with
layered structures}

\begin{figure}[h!]
\begin{center}$
\begin{array}{cccc}
\includegraphics[width=50 mm]{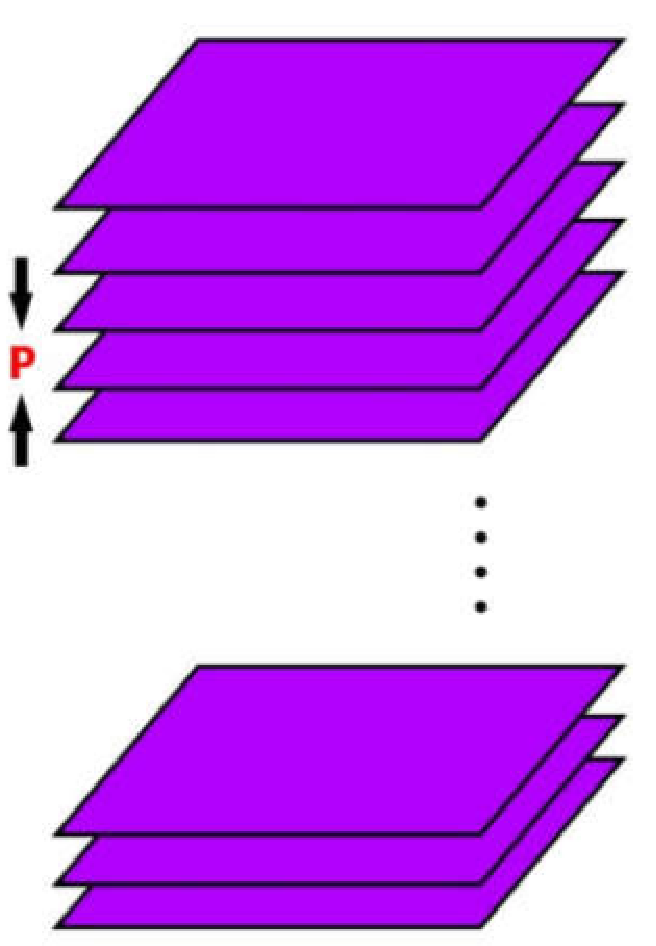}
\end{array}$
\end{center}
\caption{{\it{Atomic layers contain heavy atoms in the superlattices
of solids with layered structures that lead to extremely high
pressures.} }}\label{fig5}
\end{figure}

\begin{figure}[h!]
\begin{center}$
\begin{array}{cccc}
\includegraphics[width=60 mm]{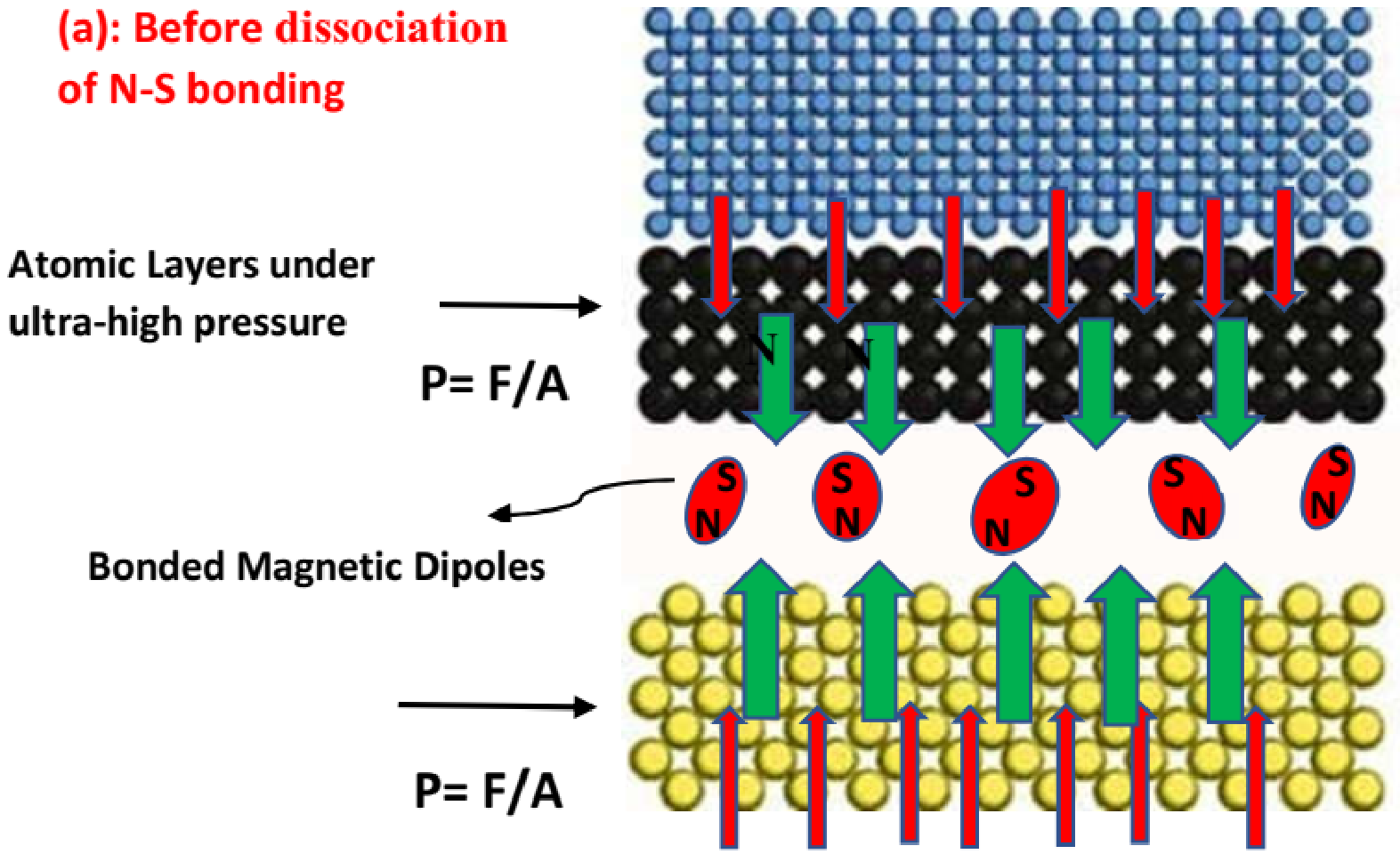}\includegraphics[width=60 mm]{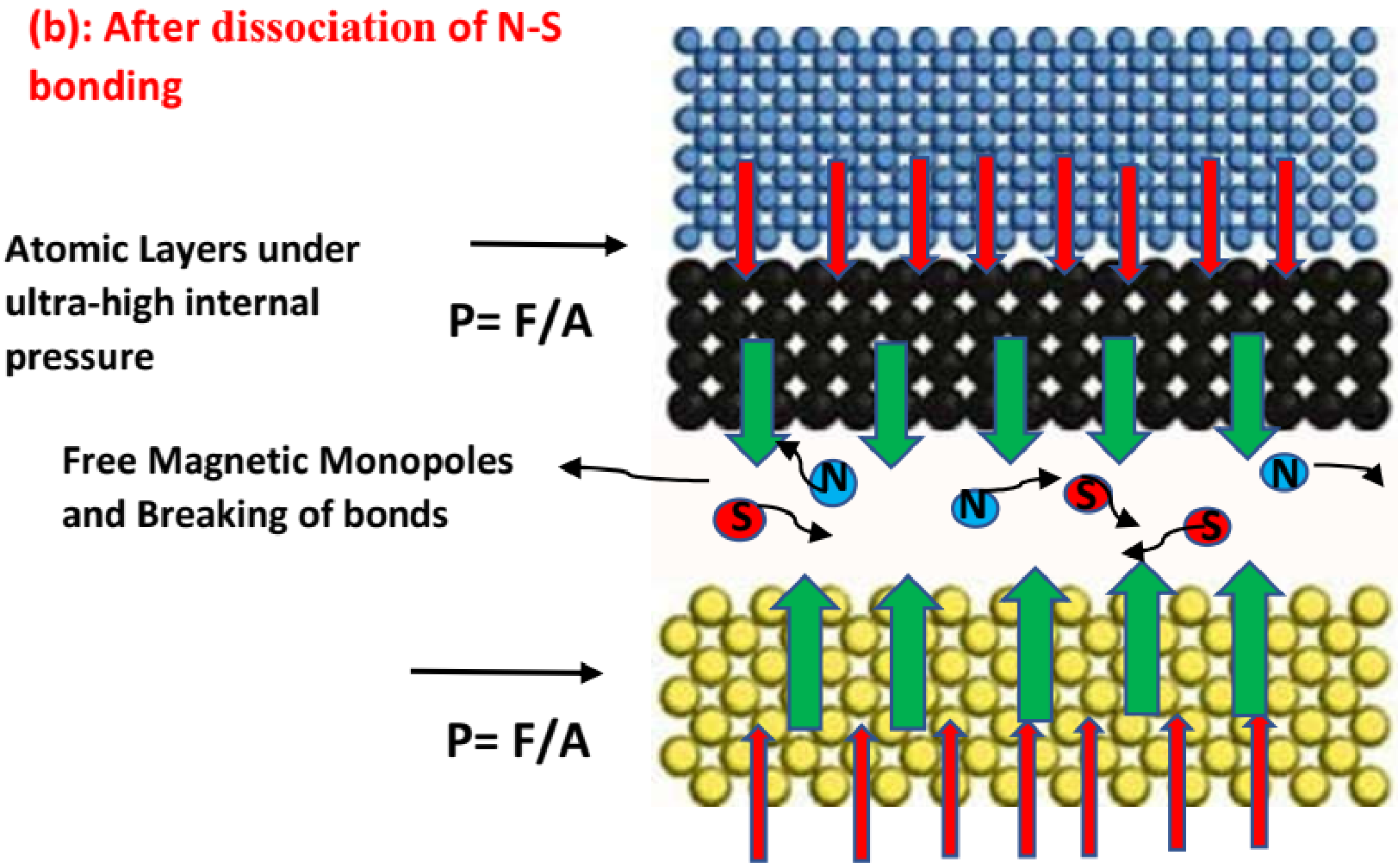}
\end{array}$
\end{center}
\caption{{\it{(a) Magnetic dipoles under ultra-high pressure between
atomic layers before bond breaking. (b) Free magnetic monopoles
between atomic layers after bond breaking and mobility of free
monopoles.}}} \label{fig6}
\end{figure}

Let us now proceed to a different potential  scenario for the separation of
magnetic poles is in a superlattice of layered nanostructured solids, where
extremely high pressures are created between the plates. In fact, in the
presence of heavy ions, the ultra-high pressures in these layered solids may
lead to the decomposition of magnetic poles (as shown in Fig. \ref{fig5}).
The pressure  between the plates in a solid-state superlattice can be
calculated using the P-E relationships in solid-state physics. However,
according to thermodynamics, the relationships between internal pressure ($P$)
and energy density ($E$) in solids are given by
\begin{equation}\label{17}
P=-\left(\frac{\partial E}{\partial V}\right)_{N},
\end{equation}
and
\begin{equation}\label{18}
P=\frac{2}{3}\frac{E}{V},
\end{equation}
where $E$ and $V$ represent the energy density and volume of the solid,
respectively. Using these equations, we can estimate the pressure between the
plates in a solid-state superlattice, given the energy value from
(\ref{15}). If we assume $V= a^3$, where $a$ is the lattice constant (equal to
$5\times10^{-10} m$ or 5 ${\r{A}}$  for a typical solid lattice), then for an
energy of $E \approx10^{-16} J$, the pressure ($P$) is on the order of
$\approx10^{12}$ Pascal. Such high pressures are achievable between the plates
of a superlattice. The pressure between the plates in a solid-state superlattice
can be estimated using thermodynamic equations, and can reach extremely high
values due to the ultra-high pressures created between the plates. In Fig.
\ref{fig6} we illustrate the formation of magnetic dipoles under ultra-high
pressure between atomic layers in a solid-state superlattice, both before bond
breaking (Fig. \ref{fig6}(a)) and after the formation of free magnetic
monopoles (Fig. \ref{fig6}(b)). The N-S dipoles between the atomic layers
convert into two single charges after bond breaking, resulting to the
production of a super-current due to the mobility of the free magnetic
monopoles.

The production of  a super-current in a solid-state superlattice with free
magnetic monopoles has important practical implications. Super-currents are a
type of electrical current that flow with zero resistance, which means that they
can flow indefinitely without any loss of energy. This property is known as
superconductivity, and it is observed in certain materials at very low
temperatures. The production of a super-current due to the mobility of free
magnetic monopoles in a solid-state superlattice is significant because it
suggests the possibility of creating new types of superconducting materials that
operate at higher temperatures and under less extreme conditions. Currently,
most superconducting materials require extremely low temperatures and high
pressures to achieve superconductivity, which limits their practical
applications. If a solid-state superlattice could be engineered to produce a
super-current at higher temperatures and pressures, it could have important
applications in a wide range of fields, such as energy transmission, magnetic
levitation, and high-speed computing. Additionally, the discovery of new types
of superconducting materials could have a profound impact on our understanding
of condensed matter physics and could lead to new discoveries in fundamental
physics.

\section{Variational method}

The variational method is a powerful and versatile tool in the analysis of
magnetic dipoles and their interactions with other particles \cite{42}. It is a
mathematical technique that allows us to estimate the ground state energy of a
quantum mechanical system by choosing a trial wave function that approximates
the ground state wave function of the system. The energy of the trial wave
function is then calculated and minimized with respect to the parameters of the
wave function, yielding an estimate of the ground state energy of the system.

In the context of magnetic dipoles, the variational  method has been used to
estimate the energy of a magnetic dipole in the presence of a magnetic field.
This can provide valuable insights into the behavior of magnetic dipoles and
their interactions with other particles, and can help to guide the design of new
materials and devices. Additionally, it has  been applied in a wide range of
studies related to magnetic dipoles. For instance, it has been used to
investigate the properties of magnetic nanoparticles, which are important for a
variety of applications such as data storage, biomedical imaging, and drug
delivery.

The variational method has been  used  to estimate the energy of a magnetic
nanoparticle in the presence of an external magnetic field \cite{43}. The trial wave
function is chosen to be a linear combination of atomic orbitals, and the
energy is minimized with respect to the coefficients of the wave function.
Hence, the resulting energy estimate is used to study the magnetic properties
of the particle and its interactions with other particles.

Another example of the variational method application in the analysis of
magnetic dipoles, is in the study of spin crossover compounds. Spin crossover
compounds are molecules that can switch between two different spin states, and
are of interest for their potential applications in data storage and molecular
electronics. Additionally, the variational method was used to estimate the
energy of a spin crossover compound in the presence of a magnetic field. The
trial wave function was chosen to be a linear combination of atomic orbitals,
and the energy was minimized with respect to the coefficients of the wave
function. The resulting energy estimation was used to study the properties of
the compound in different spin states, and to explore its potential applications
in data storage and molecular electronics.

Let us summarize here some  examples of the variational method application in
magnetic dipole analysis:\\
1- Study of magnetic domain walls: Domain walls are interfaces between regions
of different magnetic orientations and are important in magnetic data storage
and spintronics. The variational method has been used to estimate the energy of
a domain wall in the presence of a magnetic field, and to study the properties
of the wall and its interactions with other particles.\\
2- Calculation of magnetic moments: The magnetic moment is a fundamental
property of magnetic dipoles and is critical for understanding their behavior
and interactions. The variational method has been used to estimate the magnetic
moment of a dipole in the presence of a magnetic field, and to study how the
moment changes with different external conditions.\\
3- Analysis of magnetic anisotropy: Magnetic anisotropy refers to the
directional dependence of the magnetic properties of a material, and is
important in a variety of applications such as data storage and magnetic
sensing. The   method has been used to estimate the energy of a
magnetic dipole in the presence of an anisotropic magnetic field, and to study
the anisotropy of different materials.\\
4- Investigation of magnetic nanoparticles: Magnetic nanoparticles are
important for a variety of applications such as biomedical imaging and drug
delivery. The   method has been used to estimate the energy of a nanoparticle in
the presence of an external magnetic field, and to study the magnetic properties
of the particle and its interactions with other particles.
Finally, the   method has been used in the design and
development of new magnetic materials with specific properties. One example is
its  use to optimize the magnetic properties of magnetic alloys.

While the  variational method  is a powerful tool in the design and
optimization of magnetic materials, it is not without limitations and
challenges. Here we present some of the limitations and challenges of using the
  method in material design:\\
1- Choice of trial wave function: The accuracy of the  method depends
on the choice of the trial wave function. If the trial wave function is not a
good approximation of the ground state wave function, the energy estimation
will be inaccurate. Choosing an appropriate trial wave function can be
challenging, especially for complex materials.\\
2- Computational complexity: The variational method involves the computation of
integrals and the minimization of energy with respect to the parameters of the
wave function. These calculations can be computationally intensive and require
significant computational resources, especially for larger and more complex
systems.\\
3- Sensitivity to external conditions: The energy estimates obtained from the
variational method can be sensitive to external conditions such as temperature,
pressure, and magnetic field strength. This can make it difficult to accurately
predict the behavior of a material under different conditions.\\
4- Limited applicability: The  method is a quantum mechanical
technique and is therefore limited to the analysis of materials at the atomic
and molecular level. It may not be applicable to the analysis of materials at
larger scales.

However, despite the aforementioned  limitations and challenges, the
variational method remains a valuable tool in the design and optimization of
magnetic materials. Its ability to estimate the energy of a quantum mechanical
system in the presence of external fields makes it a powerful technique for
optimizing the magnetic properties of materials for specific applications.

\section{Conclusions}

In this work, we investigated the conditions under which magnetic dipoles can
separate. As we saw, we found that in a classical model of a harmonic
oscillator, isolated magnetic monopoles exist in the presence of very
high energy or temperature and pressure. A similar result was obtained using
quantum mechanics at   a good approximation level. Our results suggest that the
separation of magnetic poles, as the fifth phase of matter, can occur in two
states: (a) in a very hot plasma environment with extremely high temperatures,
such as in the center of a hot star, and (b) at extremely high pressures, such
is the case between internal plates in complex superlattices of layered
solids.

In particular, we showed that the simple harmonic oscillator model, in both
classical and quantum mechanics, is successful in calculating the breaking
energy of a magnetic dipole particle. Moreover, the model suggests that it
is possible to create magnetic monopoles in hot plasma media, with a breakdown
temperature ($\theta$) on the order of $10^7$,  results  that are
consistent in both classical and quantum approaches. Furthermore,
 based on this model, we   calculated that the dissociation of bonds
between N and S magnetic poles in solid superlattices occurs at very high
pressures between crystal plates. For a typical solid superlattice, if the
energy is $E\approx10^{-16} J$ then the pressure can be on the order of
$P\approx10^{12}$ Pascal, a value that is achievable between
plates in superlattices such as perovskite and pyrochlore. Lastly, the model
may have potential practical applications in constructing new spin devices and
advanced magnetic-electronics materials with magnetic monopole carriers using
layered superlattice solids.

\section*{Acknowledgments}

P.R. acknowledges the Inter University Centre for Astronomy and
Astrophysics (IUCAA), Pune, India for granting visiting
associateship.

\end{document}